\PassOptionsToPackage{unicode}{hyperref}
\PassOptionsToPackage{hyphens}{url}
\documentclass[
]{article}
\usepackage{amsmath,amssymb}
\usepackage{iftex}
\ifPDFTeX
  \usepackage[T1]{fontenc}
  \usepackage[utf8]{inputenc}
  \usepackage{textcomp} % provide euro and other symbols
\else % if luatex or xetex
  \usepackage{unicode-math} % this also loads fontspec
  \defaultfontfeatures{Scale=MatchLowercase}
  \defaultfontfeatures[\rmfamily]{Ligatures=TeX,Scale=1}
\fi
\usepackage{lmodern}
\ifPDFTeX\else
\fi
\IfFileExists{upquote.sty}{\usepackage{upquote}}{}
\IfFileExists{microtype.sty}{% use microtype if available
  \usepackage[]{microtype}
  \UseMicrotypeSet[protrusion]{basicmath} % disable protrusion for tt fonts
}{}
\makeatletter
\@ifundefined{KOMAClassName}{% if non-KOMA class
  \IfFileExists{parskip.sty}{%
    \usepackage{parskip}
  }{% else
    \setlength{\parindent}{0pt}
    \setlength{\parskip}{6pt plus 2pt minus 1pt}}
}{% if KOMA class
  \KOMAoptions{parskip=half}}
\makeatother
\usepackage{xcolor}
\usepackage[margin=1in]{geometry}
\usepackage{color}
\usepackage{fancyvrb}

\DefineVerbatimEnvironment{Highlighting}{Verbatim}{commandchars=\\\{\}}
\newenvironment{Shaded}{}{}

\newcommand{\AttributeTok}[1]{\textcolor[rgb]{0.49,0.56,0.16}{#1}}

\newcommand{\BuiltInTok}[1]{\textcolor[rgb]{0.00,0.50,0.00}{#1}}

\newcommand{\CommentTok}[1]{\textcolor[rgb]{0.38,0.63,0.69}{\textit{#1}}}

\newcommand{\ControlFlowTok}[1]{\textcolor[rgb]{0.00,0.44,0.13}{\textbf{#1}}}

\newcommand{\DecValTok}[1]{\textcolor[rgb]{0.25,0.63,0.44}{#1}}

\newcommand{\ImportTok}[1]{\textcolor[rgb]{0.00,0.50,0.00}{\textbf{#1}}}

\newcommand{\KeywordTok}[1]{\textcolor[rgb]{0.00,0.44,0.13}{\textbf{#1}}}
\newcommand{\NormalTok}[1]{#1}
\newcommand{\OperatorTok}[1]{\textcolor[rgb]{0.40,0.40,0.40}{#1}}

\newcommand{\StringTok}[1]{\textcolor[rgb]{0.25,0.44,0.63}{#1}}

\usepackage{graphicx}
\makeatletter
\def\maxwidth{\ifdim\Gin@nat@width>\linewidth\linewidth\else\Gin@nat@width\fi}
\def\maxheight{\ifdim\Gin@nat@height>\textheight\textheight\else\Gin@nat@height\fi}
\makeatother
\setkeys{Gin}{width=\maxwidth,height=\maxheight,keepaspectratio}
\makeatletter
\def\fps@figure{htbp}
\makeatother
\ifLuaTeX
  \usepackage{selnolig}  % disable illegal ligatures
\fi
\usepackage[]{biblatex}
\IfFileExists{bookmark.sty}{\usepackage{bookmark}}{\usepackage{hyperref}}
\IfFileExists{xurl.sty}{\usepackage{xurl}}{} % add URL line breaks if available
\hypersetup{
  pdftitle={PARMESAN: Meteorological Timeseries and Turbulence Analysis Backed by Symbolic Mathematics},
  pdfauthor={Yann Georg Büchau1,; Hasan Mashni1; Matteo Bramati1; Vasileios Savvakis1; Ines Schäfer1; Saskia Jung1; Gabriela Miranda-Garcia1; Daniel Hardt2; Jens Bange1},
  hidelinks,
  pdfcreator={LaTeX via pandoc}}

\title{PARMESAN: Meteorological Timeseries and Turbulence Analysis
Backed by Symbolic Mathematics}
\author{Yann Georg Büchau\textsuperscript{1,*} \and Hasan Mashni\textsuperscript{1} \and Matteo Bramati\textsuperscript{1} \and Vasileios Savvakis\textsuperscript{1} \and Ines Schäfer\textsuperscript{1} \and Saskia Jung\textsuperscript{1} \and Gabriela Miranda-Garcia\textsuperscript{1} \and Daniel Hardt\textsuperscript{2} \and Jens Bange\textsuperscript{1}}
\date{26 September 2023}

\begin{document}
\maketitle

\textsuperscript{1} Eberhard Karls Universität Tübingen, Germany\\
\textsuperscript{2} Akaflieg Braunschweig e.V., Braunschweig, Germany

\textsuperscript{*} Correspondence:
\href{mailto:yann-georg.buechau@uni-tuebingen.de}{Yann Georg Büchau
\textless{}yann-georg.buechau@uni-tuebingen.de\textgreater{}}

\hypertarget{summary}{%
\section{Summary}\label{summary}}

PARMESAN (the \textbf{P}ython \textbf{A}tmospheric \textbf{R}esearch
Package for \textbf{ME}teorological Time\textbf{S}eries and Turbulence
\textbf{AN}alysis) is a Python package providing common functionality
for atmospheric scientists doing time series or turbulence analysis.
Several meteorological quantities such as potential temperature, various
humidity measures, gas concentrations, wind speed and direction,
turbulence and stability parameters can be calculated. Furthermore,
signal processing functionality such as properly normed variance spectra
for frequency analysis is available. In contrast to existing packages
with similar goals, its routines for physical quantities are derived
from symbolic mathematical expressions, enabling inspection, automatic
rearrangement, reuse and recombination of the underlying equations.
Building on this, PARMESAN's functions as well as their comprehensive
parameter documentation are mostly auto-generated, minimizing human
error and effort. In addition, sensitivity/error propagation analysis is
possible as mathematical operations like derivations can be applied to
the underlying equations. Physical consistency in terms of units and
value domains are transparently ensured for PARMESAN functions.
PARMESAN's approach can be reused to simplify implementation of robust
routines in other fields of physics.

\hypertarget{statement-of-need}{%
\section{Statement of need}\label{statement-of-need}}

The need to assert properly balanced physical units right from within
running programs and models has been recognised for a long time now
\autocite{cooper2008,chizeck2009errordetectionunit}. Unit conversion
errors in science and engineering have caused costly system failures
such as the NASA Mars Climate Orbiter crash in 1999
\autocite{nasa1999marsclimateorbiter}.

Nowadays, the Python ecosystem comprises many packages that ease
specific tasks when performing physical calculations: \texttt{numpy}
\autocite{numpy} and \texttt{scipy} \autocite{scipy} provide efficient
numerical routines, \texttt{pandas} \autocite{pandas} and
\texttt{xarray} \autocite{xarray} provide structures to read, write and
aggregate data, \texttt{pint} \autocite{pint} handles physical units and
the \texttt{uncertainties} package \autocite{uncertainties} simplifies
linear error propagation. Partly based on those, collections of routines
for atmospheric science exist such as \texttt{metpy} \autocite{metpy},
\texttt{iris} \autocite{Iris} and \texttt{aoslib}/PyAOS
\autocite{PyAOS}. However these focus more on gridded, spatial data
which is common in modelling and remote sensing and have little
functionality for turbulence analysis. Turbulence plays an important
role in atmospheric exchange processes, especially in the planetary
boundary layer \autocite{stull1988introductionboundarylayer}. It is a
statistical process and thus mostly quantified through high-resolution
in-situ measurement techniques
\autocite{foken2021springerhandbookatmospheric}. \texttt{metpy} and
\texttt{iris} can both handle units and require the user to explicitly
specify them. Their physical quantities are calculated using hard-coded
expressions. In contrast, the \texttt{atmos} package \autocite{atmos}
has implemented an equation solving system for more flexible reusability
and less hard-coding of relationships between quantities. Its
development seems to have stalled since 2020, though. None of the above
packages have a mechanism for transparently checking that input and
output values are within reasonable physical bounds.

PARMESAN addresses the aforementioned gaps by providing functions for
meteorological quantities that are backed by symbolic mathematical
expressions employing SymPy \autocite{sympy}, a powerful computer
algebra system written purely in Python. Inputs and outputs are checked
for and potentially converted to correct units while asserting that the
physical domains are not exceeded. It can rearrange its equations and
thus flexibly increase the number of available functions. PARMESAN has
already been used successfully in
\textcite{buchau2022autarkicwirelesssensor} for data analysis of
atmospheric measurements.

\hypertarget{structure}{%
\section{Structure}\label{structure}}

\begin{figure}
\centering
\includegraphics{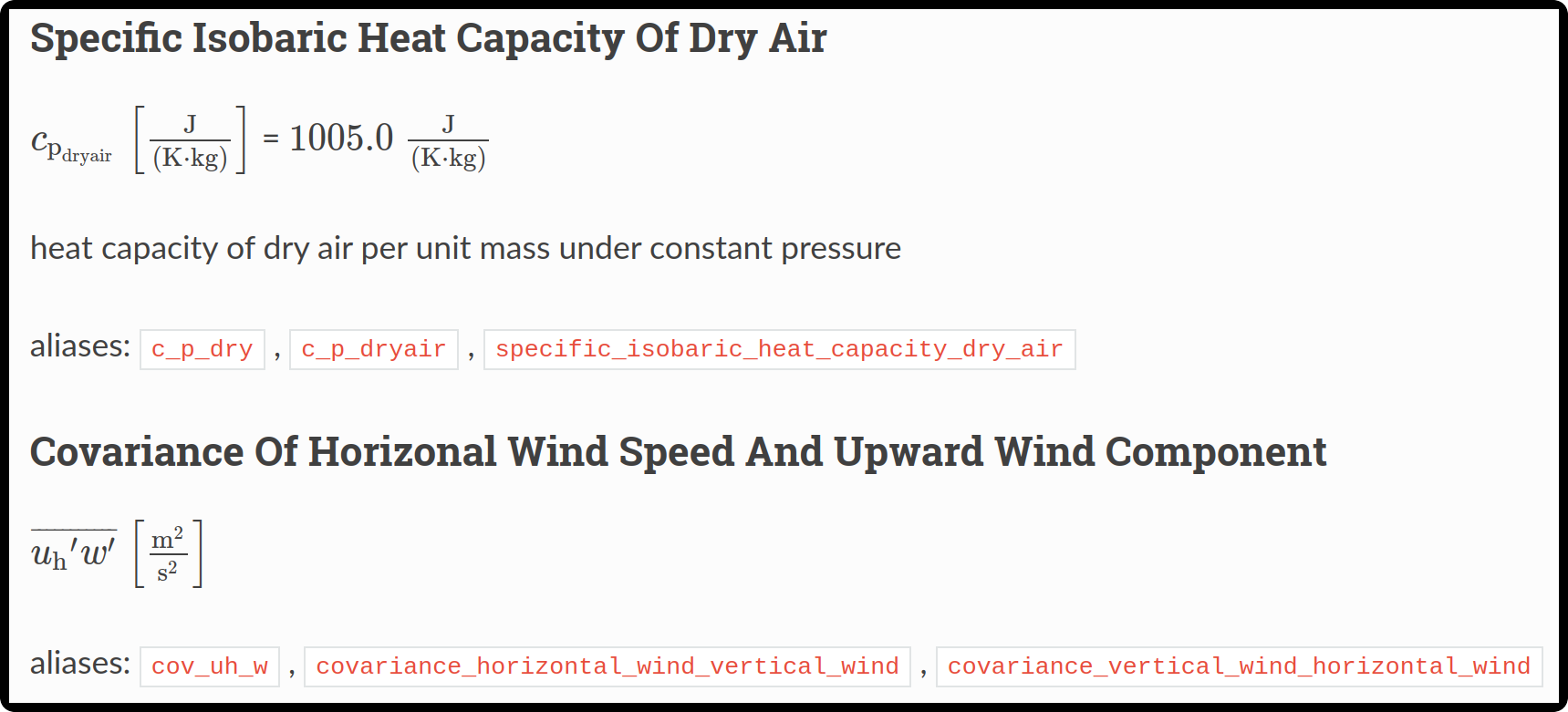}
\caption{Excerpt of auto-generated symbol list in
\texttt{parmesan.symbols}. Symbols have metadata such as descriptions,
units and default values attached. For readability, they can be referred
to with different variable names, which are also available as parameter
aliases when calling functions in PARMESAN. \label{FIGxSymbolList}}
\end{figure}

Functions for physical quantities in PARMESAN are based on symbolic
mathematical equations created using SymPy \autocite{sympy}. PARMESAN
defines a descriptive list of symbols (i.e.~variables and constants,
\autoref{FIGxSymbolList}) and relates them to form the common laws of
thermodynamics, parametrisations and definitions used in atmospheric
science.

\begin{figure}
\centering
\includegraphics{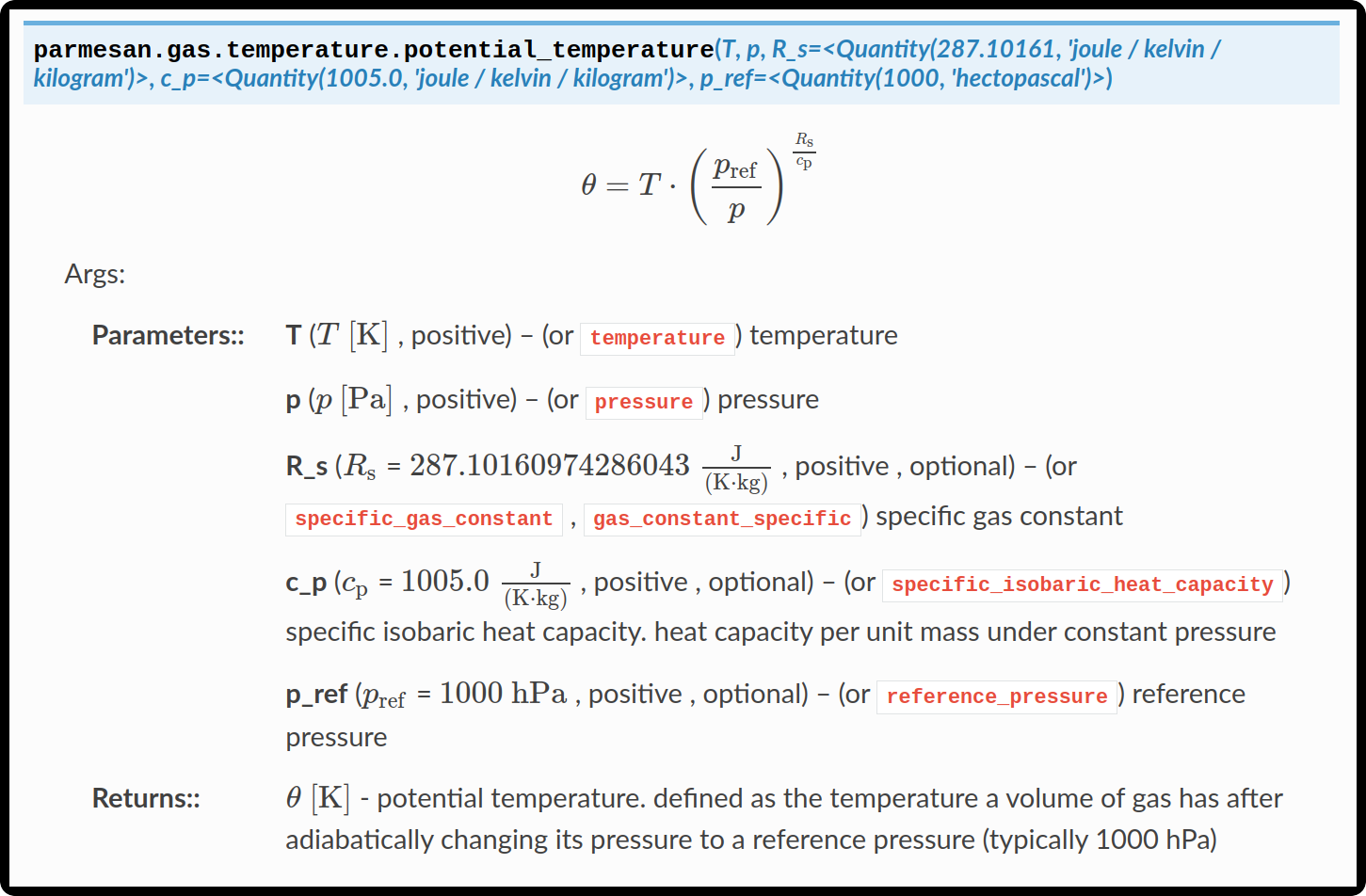}
\caption{Auto-generated comprehensive parameter documentation and
LaTeX-formatted equation for PARMESAN's
\texttt{potential\_temperature()} function to calculate potential
temperature from atmospheric pressure and temperature. Parameter
aliases, units, defaults and bounds are taken from PARMESAN's symbol
library (\autoref{FIGxSymbolList}) and used coherently across functions
in PARMESAN. \label{FIGxPotentialTemperatureDocs}}
\end{figure}

This approach has many advantages over the traditional method of
hard-coding mathematical operations between function inputs using
language-specific constructs. First of all, information about the
mathematical relationship between quantities is not lost, but can
instead be queried and reused. SymPy equations can be rearranged and
recombined to generate new expressions, enabling the generation of many
specific functions from a set of base equations. Additionally, SymPy
expressions are translatable into code for numerous programming
languages. PARMESAN uses this mechanism to turn its equations into
executable Python functions that use the efficient \texttt{numpy}
package internally \autocite{numpy}, so no runtime overhead is
introduced and array inputs and outputs are supported. Symbolic
expressions are automatically simplified and terms cancelled
accordingly, revealing the set of input parameters an equation really
depends on. This information is then used to automatically generate
extensive documentation for each individual function
(\autoref{FIGxPotentialTemperatureDocs}) - a great benefit for
consistency and minimisation of human effort and oversight in the
documentation.

\pagebreak

Creating new functions in or from PARMESAN thus often requires only very
few lines of code. Here is a compacted version of PARMESAN's function
for potential temperature specifically for dry air:

\begin{Shaded}
\begin{Highlighting}[]
\ImportTok{from}\NormalTok{ parmesan.symbols }\ImportTok{import} \OperatorTok{*}   \CommentTok{\# Import all of PARMESAN\textquotesingle{}s symbols}
\AttributeTok{@from\_sympy}\NormalTok{() }\CommentTok{\# decorator turning SymPy expression into code and documentation}
\KeywordTok{def}\NormalTok{ potential\_temperature(): }\CommentTok{\# no arguments necessary, added automatically}
    \ControlFlowTok{return}\NormalTok{ T }\OperatorTok{*}\NormalTok{ (p\_ref }\OperatorTok{/}\NormalTok{ p) }\OperatorTok{**}\NormalTok{ (R\_dryair }\OperatorTok{/}\NormalTok{ c\_p\_dryair)   }
    \CommentTok{\# SymPy expression {-} practically equal to typical Python code}
\end{Highlighting}
\end{Shaded}

In this case, the resulting quantity is derived from the function's
name, documentation is generated
(\autoref{FIGxPotentialTemperatureDocs}) and the equation is immediately
checked for units consistency employing the \texttt{pint} package
\autocite{pint}. Each symbol has metadata attached, such as a physical
unit and a domain (\autoref{FIGxSymbolList}). These are available to the
resulting function for assertion, so a PARMESAN function will check and
auto-convert input and output units and issue a warning when unphysical
values arise such as negative absolute temperatures:

\begin{Shaded}
\begin{Highlighting}[]
\CommentTok{\# Implicit Units }
\NormalTok{potential\_temperature(T}\OperatorTok{=}\DecValTok{300}\NormalTok{, p}\OperatorTok{=}\DecValTok{100000}\NormalTok{) }\CommentTok{\# K and Pa assumed}
\CommentTok{\# 300.0 K}

\CommentTok{\# Explicit Units}
\ImportTok{from}\NormalTok{ parmesan.units }\ImportTok{import}\NormalTok{ units }\CommentTok{\# PARMESAN\textquotesingle{}s predefined units}
\NormalTok{potential\_temperature(T}\OperatorTok{=}\NormalTok{units.Quantity(}\DecValTok{20}\NormalTok{,}\StringTok{"°C"}\NormalTok{), p}\OperatorTok{=}\DecValTok{950} \OperatorTok{*}\NormalTok{ units.hPa)}
\CommentTok{\# 297.477188635086 K}

\CommentTok{\# Parameter/Symbol Aliases}
\NormalTok{potential\_temperature(temperature}\OperatorTok{=}\DecValTok{300}\NormalTok{, pressure}\OperatorTok{=}\DecValTok{100000}\NormalTok{) }
\CommentTok{\# 300.0 K}

\CommentTok{\# Arrays}
\ImportTok{import}\NormalTok{ numpy }\ImportTok{as}\NormalTok{ np}
\NormalTok{potential\_temperature(T}\OperatorTok{=}\DecValTok{300}\NormalTok{, p}\OperatorTok{=}\NormalTok{np.array([}\DecValTok{950}\NormalTok{,}\DecValTok{980}\NormalTok{,}\DecValTok{1010}\NormalTok{]) }\OperatorTok{*}\NormalTok{ units.hPa)}
\CommentTok{\# Magnitude: [304.42830151978785 301.7364178157801 299.14844787358106]}
\CommentTok{\#     Units: K}

\CommentTok{\# Bounds check}
\NormalTok{potential\_temperature(T}\OperatorTok{={-}}\DecValTok{10}\NormalTok{, p}\OperatorTok{=}\DecValTok{1010}\OperatorTok{*}\NormalTok{units.hPa) }\CommentTok{\# temperature out of bounds}
\CommentTok{\# OutOfBoundsWarning: 1 of 1 input values to potential\_temperature for }
\CommentTok{\# argument \textquotesingle{}T\textquotesingle{} are out of bounds defined by \textquotesingle{}positive\textquotesingle{}: [{-}10] at indices [0]}
\CommentTok{\# {-}9.971614929119369 K}

\CommentTok{\# Units check}
\NormalTok{potential\_temperature(T}\OperatorTok{=}\DecValTok{300}\NormalTok{, p}\OperatorTok{=}\DecValTok{1010}\OperatorTok{*}\NormalTok{units.degrees) }\CommentTok{\# wrong unit {-}\textgreater{} error}
\CommentTok{\# DimensionalityError: potential\_temperature(): }
\CommentTok{\#    p=\textless{}Quantity(1010, \textquotesingle{}degree\textquotesingle{})\textgreater{} could not be converted to pascal: }
\CommentTok{\#    Cannot convert from \textquotesingle{}degree\textquotesingle{} (dimensionless) to \textquotesingle{}pascal\textquotesingle{} }
\CommentTok{\#    ([mass] / [length] / [time] ** 2)}
\end{Highlighting}
\end{Shaded}

\begin{figure}
\centering
\includegraphics{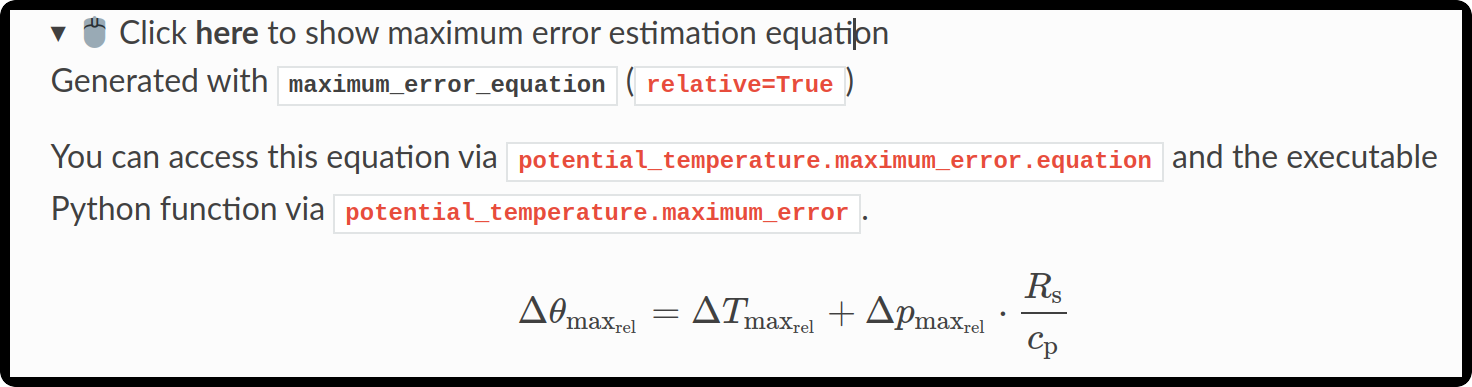}
\caption{Auto-generated maximum relative error equation
(\autoref{EQUATIONxMaximumErrorEquation}) for PARMESAN's
\texttt{potential\_temperature()} function
(\autoref{FIGxPotentialTemperatureDocs}). Symbolic PARMESAN functions
automatically have a sensitivity analysis attached to quantify how a
change in input parameters affects the output. In this case, the maximum
expected relative error of potential temperature {[}\%{]} is the sum of
the maximum relative errors of temperature and pressure {[}\%{]}, with
the pressure term scaled by a factor.
\label{FIGxPotentialTemperatureDocsMaxErr}}
\end{figure}

Another benefit of having the underlying symbolic expression for an
equation available is the possibility to do sensitivity analysis.
PARMESAN can derive the maximum relative error
\(\Delta y_\text{max,rel}\) (\autoref{EQUATIONxMaximumErrorEquation})
for its symbolic functions
(\autoref{FIGxPotentialTemperatureDocsMaxErr}):

\begin{equation}
    \begin{split}
        \Delta y_\text{max}(x_1, \dots, x_\text{n})
        &=
        \sum_{i=1}^n
        \left| \frac{\partial y}{\partial x_i} \right|
        \cdot \Delta x_{i_\text{max}}
        \\[1em]
        \Delta y_\text{max,rel}
        &=
        \frac{\Delta y_\text{max}}{\overline{y}}
    \end{split}
    \label{EQUATIONxMaximumErrorEquation}
\end{equation}

The maximum relative error is a conservative estimation method for the
propagation of errors of input quantities \(x_i\) to effective error in
the output quantity \(y\), assuming the most severe combination of input
quantity deviations \(\Delta x_{i_\text{max}}\). Custom sensitivity
analyses can also be implemented based on PARMESAN's equations.

\begin{figure}
\centering
\includegraphics{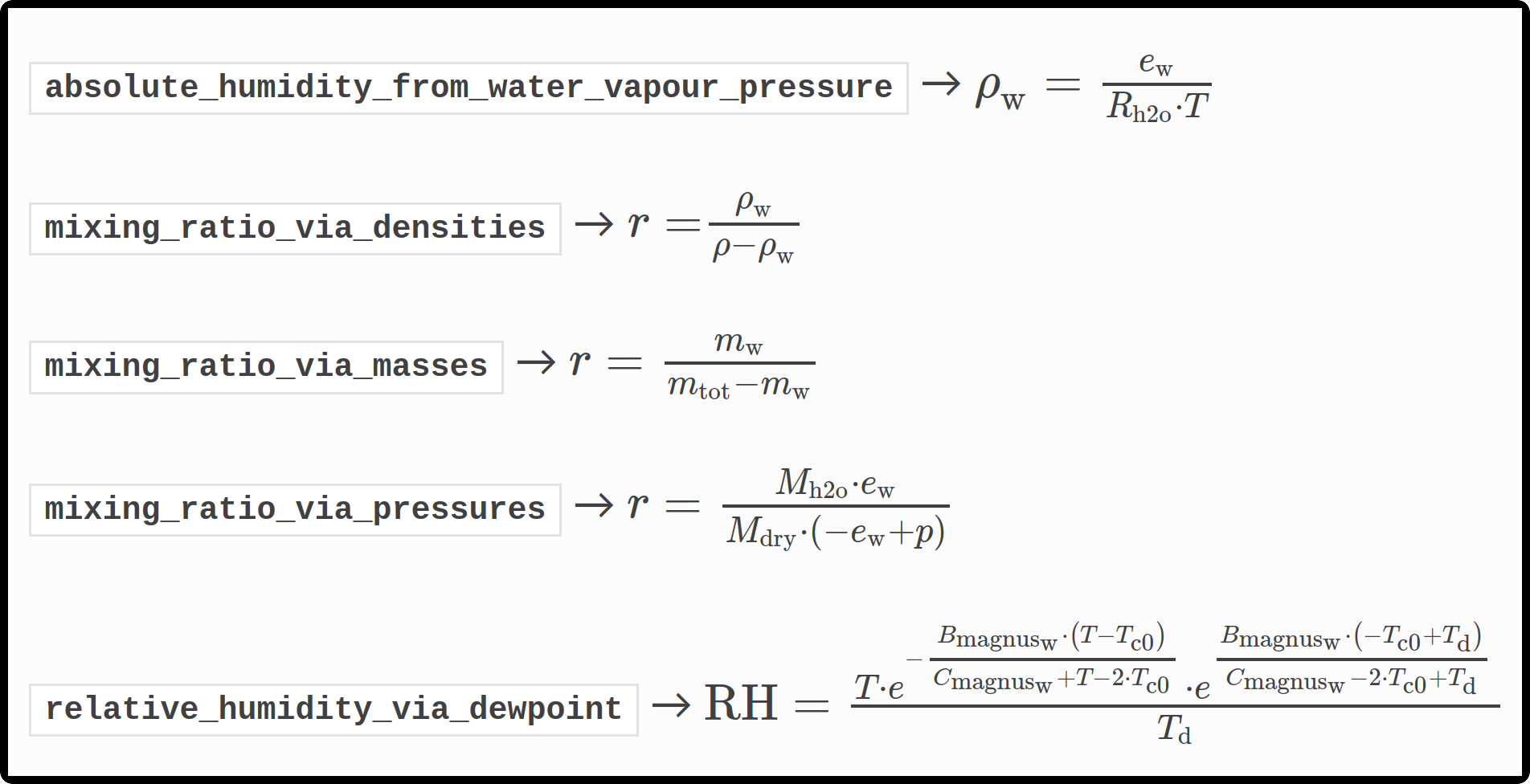}
\caption{Excerpt of auto-generated humidity equation list in PARMESAN's
\texttt{humidity} module. As the underlying equations in PARMESAN's
functions are available as symbolic expressions, it can provide
overviews of all related equations. \label{FIGxEquationList}}
\end{figure}

PARMESAN can also rearrange its existing equations
(\autoref{FIGxEquationList}) for a quantity of interest by its provided
\texttt{get\_function()} function:

\begin{Shaded}
\begin{Highlighting}[]
\ImportTok{from}\NormalTok{ parmesan.symbols }\ImportTok{import} \OperatorTok{*}
\CommentTok{\# get (or rearrange) functions that calculate mixing ratio}
\NormalTok{mixing\_ratio\_functions }\OperatorTok{=} \BuiltInTok{list}\NormalTok{(get\_function(result}\OperatorTok{=}\NormalTok{mixing\_ratio)}
\CommentTok{\# get (or rearrange) functions that calculate mixing ratio}
\CommentTok{\# from at least temperature and pressure}
\NormalTok{mixing\_ratio\_functions }\OperatorTok{=} \BuiltInTok{list}\NormalTok{(get\_function(result}\OperatorTok{=}\NormalTok{mixing\_ratio, inputs}\OperatorTok{=}\NormalTok{(T, p))}
\end{Highlighting}
\end{Shaded}

The functions found can be called as usual or their underlying equations
can be examined by accessing their \texttt{.equation} attribute. In a
Jupyter notebook \autocite{kluyver2016jupyternotebookspublishing} the
equations appear as formatted markup similar to what is depicted in
\autoref{FIGxEquationList}.

\begin{figure}
\centering
\includegraphics{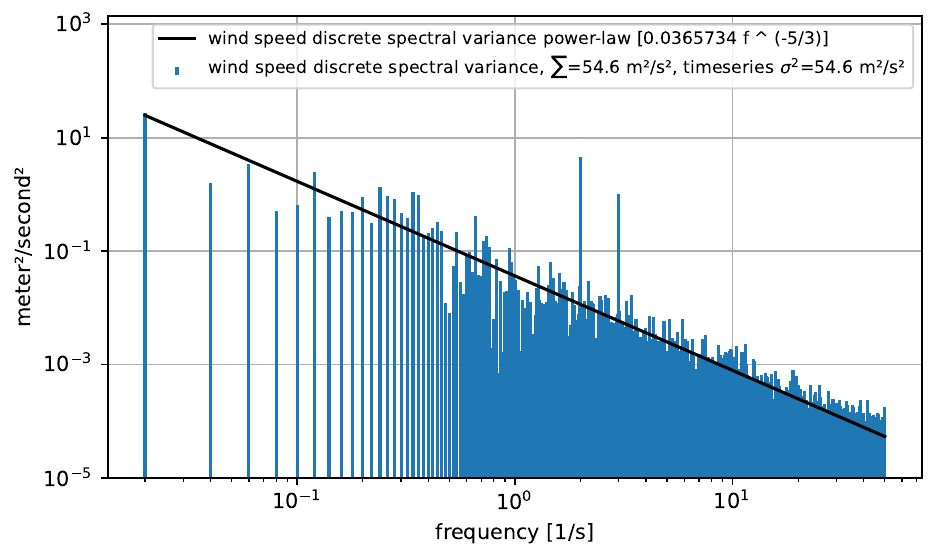}
\caption{PARMESAN discrete \texttt{variance\_spectrum()} of an
artificial wind timeseries (random walk overlayed with 2Hz and 3Hz sine
waves). Note how Parseval's Theorem
\autocite{stull1988introductionboundarylayer} is correctly fulfilled as
the timeseries variance equals the sum of discrete spectral variances. A
Kolmogorov power-law fit \autocite{ortizsuslow2019evaluationkolmogorovs}
was optionally added by PARMESAN. \label{FIGxSpectrum}}
\end{figure}

Besides physical equations, PARMESAN provides tools often needed when
analysing timeseries such as calculating second-order moments, variance
spectrum (\autoref{FIGxSpectrum}), autocorrelation, structure function
(variogram) and running covariance, e.g.~for calculating eddy fluxes
\autocite{foken2021springerhandbookatmospheric}, backed by the
\texttt{scipy} package \autocite{scipy} for efficient numerics and
\texttt{matplotlib} \autocite{matplotlib} for visualisation. PARMESAN
integrates with the common \texttt{pandas} data analysis framework
\autocite{pandas} by adding a \texttt{.parmesan} accessor to
\texttt{DataFrame} and \texttt{Series} objects to apply PARMESAN
functions such as a variance spectrum or autocorrelation directly to
them.

\hypertarget{acknowledgements}{%
\section{Acknowledgements}\label{acknowledgements}}

This software originated in a project funded by the German Research
Foundation (DFG) under grant number BA 1988/19-1.

\printbibliography[title=References]

\end{document}